%% file: HIN-24-001_temp.tex
\pdfoutput=1
\documentclass[11pt,twoside,a4paper,cmspaper,final,collab]{cms-tdr}
\def\svnVersion{160f19e}\def\svnDate{2025/09/08}\def\cmsCernNoTag{CERN-EP-2025-006}\def\cmsCernDate{\today}\def\cmsMessage{Published in Physical Review Letters as \href{http://dx.doi.org/10.1103/c9wp-5tq3}{\doi{10.1103/c9wp-5tq3}.}}
\begin{document}\cmsNoteHeader{HIN-24-001}

\newcommand {\pp}      {\ensuremath{\Pp\Pp}\xspace}
\newcommand {\pPb}     {\ensuremath{\Pp}Pb\xspace}
\newcommand {\pA}     {\ensuremath{\Pp}A\xspace}
\newcommand{\Ntroff}{\ensuremath{\text{N}^{{\text{offline}}}_{\text{track}}}\xspace}
\newcommand{\Ntroffcorr}{\ensuremath{\text{N}^{{\text{corr.}}}_{\text{track}}}\xspace}
\newcommand{\sigmaRatio}{\ensuremath{\sigma_{\PGyP{2S}} /\sigma_{\PJGy}}\xspace}
\newcommand{\yCM}{\ensuremath{y_{\mathrm{CM}}}\xspace}
\newcommand{\yLab}{\ensuremath{y}\xspace}
\newcommand{\branchFrac}{\ensuremath{\mathcal{B}}\xspace}
\newcommand{\Eff}{\ensuremath{\epsilon}\xspace}
\newcommand{\nTrackIndex}{\ensuremath{N_{\text{trk}}}\xspace}
\newcommand{\nTrackMB}{\ensuremath{\langle \Ntroffcorr \rangle_{\text{MB}}}\xspace}
\newcommand{\ptTrack}{\ensuremath{\pt^{\text{track}}}\xspace}
\newcommand{\diMu}{\ensuremath{\PGmp\PGmm}\xspace}
\newcommand{\lDecay}{\ensuremath{l^{\text{3D}}}\xspace}
\newcommand{\NPass}{\ensuremath{N_{\text{pass}}}\xspace}
\newcommand{\NFail}{\ensuremath{N_{\text{fail}}}\xspace}
\newcommand{\NPrompt}{\ensuremath{N_{\text{prompt}}}\xspace}
\newcommand{\NNonPrompt}{\ensuremath{N_{\text{nonprompt}}}\xspace}
\newcommand{\PromptEff}{\ensuremath{\epsilon_{\mathrm{P}}}\xspace}
\newcommand{\NonPromptEff}{\ensuremath{\epsilon_{\mathrm{NP}}}\xspace}
\newcommand{\Lxyz}{\ensuremath{L_{xyz}}\xspace}
\ifthenelse{\boolean{cms@external}}{\providecommand{\cmsLeft}{upper\xspace}}{\providecommand{\cmsLeft}{left\xspace}}
\ifthenelse{\boolean{cms@external}}{\providecommand{\cmsRight}{lower\xspace}}{\providecommand{\cmsRight}{right\xspace}}
\newcommand{\appname}{\appendixname}

\cmsNoteHeader{HIN-24-001}
\title{Observation of the Charged-Particle Multiplicity Dependence of \texorpdfstring{\sigmaRatio}{Sigma J/Psi over Sigma Psi(2S)} in \texorpdfstring{\pPb}{pPb} Collisions at \texorpdfstring{8.16\TeV}{8.16 TeV}}
\abstract{
Bound states of charm and anticharm quarks, known as charmonia, have a rich spectroscopic structure that can be used to probe the dynamics of hadron production in high-energy hadron collisions. Here, the cross section ratio of excited $ (\PGyP{2S})$ and ground state $ (\PJGy )$ vector mesons is measured as a function of the charged-particle multiplicity in proton-lead (\pPb) collisions at a center-of-mass (CM) energy per nucleon pair of 8.16\TeV. The data corresponding to an integrated luminosity of 175\nbinv were collected using the CMS detector. The ratio is measured separately for prompt and nonprompt charmonia in the transverse momentum range $6.5 < \pt < 30\GeV$ and in four rapidity ranges spanning $-2.865 < \yCM < 1.935$. For the first time, a statistically significant multiplicity dependence of the prompt cross section ratio is observed in proton-nucleus collisions. There is no clear rapidity dependence in the ratio. The prompt measurements are compared with a theoretical model which includes interactions with nearby particles during the evolution of the system. These results provide additional constraints on hadronization models of heavy quarks in nuclear collisions. 
}

\hypersetup{
pdfauthor={CMS Collaboration},
pdftitle={Observation of the multiplicity dependence of sigma[JPsi]/sigmaPsi(2S) in pPb collisions at 8.16 TeV},%
pdfsubject={CMS},
pdfkeywords={CMS, charmonia, heavy ion, pPb, excited state, co-mover effect}} 

\maketitle 
Quantum chromodynamics (QCD) is the theory describing the strong interaction between fundamental particles having color charge, quarks and gluons.
The study of QCD and its emergent properties, particularly at high energy and/or parton density, is a major goal of nuclear physics~\cite{Busza:2018rrf,CMS:2024krd,ALICE:2022wpn}. 
Bound states of heavy quark-antiquark pairs, referred to as quarkonia, are key experimental tools in this effort since they are massive enough to enable the calculation of related observables through perturbative methods, even at low momenta~\cite{PhysRevD.55.5853}.

Quarkonia production is affected by the presence of the hot quark-gluon plasma (QGP) created in relativistic heavy ion collisions~\cite{Matsui:1986dk,Karsch:1987pv,Kharzeev:1994pz,Laine:2006ns,Brambilla:2008cx,Brambilla:2010vq,Blaizot:2015hya}. The production time of heavy-flavor quarks is shorter than the QGP formation time, 
making them excellent probes of the QGP because they experience the entire evolution of the system~\cite{vanHees:2005wb}. At energies achieved by the CERN LHC, the primary influences on quarkonium 
production within the QGP are medium color screening, leading to quarkonium dissociation, along with compensating effects that cause bound state regeneration~\cite{Brambilla:2021wkt,Yao:2020xzw,Du:2017qkv,Thews:2000rj,Braun-Munzinger:2000csl,Brambilla:2016wgg,Emerick:2011xu}. 
Furthermore, excited quarkonia states respond differently to these effects as compared to more tightly bound ground states.  

Small collision systems, such as proton-proton (\pp) and proton-ion (\pA), provide baselines against which ion-ion collisions can be compared. Measurements of these small systems performed at the CERN 
SPS~\cite{NA38:1994yau,NA50:2006rdp,NA50:2003fvu,NA51:1998uun,NA60:2010wey}, FNAL~\cite{Alde:1990wa,NuSea:1999mrl,SeaQuest:2024qdw}, HERA~\cite{HERA-B:2008ymp}, the BNL RHIC~\cite{PHENIX:2019brm,PHENIX:2022nrm,STAR:2021zvb,PHENIX:2024dqs}, and the LHC~\cite{CMS:2014vjg, CMS:2016wgo, CMS:2022wfi, CMS:2018gbb, CMS:2013jsu, CMS:2020fae, LHCb:2018psc, LHCb:2016vqr, Fronze:2016gsr, ALICE:2020vjy, ATLAS:2017prf, LHCb:2017ygo} revealed a notable suppression of excited charmonia $(\PGyP{2S})$ and bottomonia
$(\PGUP{2S}, \PGUP{3S})$ mesons compared to their ground states \PJGy and \PGUP{1S}. The observed suppression cannot be wholly attributed to the presence of initial-state effects related to nuclear parton distribution functions (nPDFs)~\cite{AtashbarTehrani:2017mzi} or energy losses caused by multiparticle 
scattering in cold nuclear matter~\cite{Arleo:2014oha, Lourenco:2008sk}. Consequently, these differences in ground and excited state production motivate measurements that specifically probe the potential presence of final-state effects, such as the hypothesized formation of a small amount of QGP~\cite{Nagle:2018nvi,Grosse-Oetringhaus:2024bwr,PHENIX:2018lia}, in small collision systems.  

The comover effect~\cite{Ferreiro:2012rq,Ferreiro:2014bia} is an alternative final-state mechanism that may suppress the observed yields of excited quarkonium states in small systems without the need to generate a QGP medium. In this paradigm, comoving particles that travel along with the quark-antiquark pair can interact with nascent quarkonia states, leading to their dissociation. 
Excited states, with larger radii and smaller binding energies, are more susceptible to dissociation upon interaction with co-moving particles, resulting in relative suppression of their yields compared to ground state yields. The strength of this suppression 
is predicted to be correlated with the number of comoving particles and therefore, with the charged-particle multiplicity of the event. In \pA collisions, more particles are created at rapidities (\yLab) closer to the outgoing ion direction, resulting in a larger predicted comover suppression for quarkonia at these rapidities as compared to those moving in the proton-going direction. This asymmetry becomes more pronounced as the charged-particle multiplicity of the event increases~\cite{ATLAS:2016xpn}. 

Transport models involving the formation of a small amount of QGP~\cite{Du:2018wsj} and comover models~\cite{Ferreiro:2014bia} have successfully explained aspects of the rapidity and multiplicity dependence of quarkonia suppression in various collision systems~\cite{Gavin:1996yd, Ferreiro:2018wbd, LHCb:2023xie, ALICE:2022gpu, LHCb:2024taa}. However, large experimental uncertainties have so far prevented definitive conclusions regarding final-state effects in the charmonia sector for \pA collisions. 

In this Letter, we present a measurement of the production cross section ratio of \PGyP{2S} to \PJGy mesons in \pPb collisions at a center-of-mass (CM) energy per nucleon pair of 8.16\TeV, with proton and lead beam energies of 6.50 and 2.56\TeV per nucleon, respectively. 
The data corresponding to an integrated luminosity of 175\nbinv were collected using the CMS detector in 2016. Ground and excited charmonium vector meson states \PJGy and \PGyP{2S} are reconstructed using the dimuon decay channel. 
The ratio is measured for both prompt (directly produced via the hadronization of a quark-antiquark pair from the earliest stages of the collision) and nonprompt (originating from \Pb hadron decays) charmonia. The precision achieved in this analysis allows for quantitative assessment of the charmonium rapidity and charged-particle multiplicity dependence, enabling stringent tests of hadronization models.

The cross section for a charmonium state having a transverse momentum, \pt, and rapidity, \yLab, can be measured using the relation $\sigma (\pt, \yLab )=N (\pt, \yLab )/ [\lumi \Eff (\pt, \yLab ) \branchFrac ]$, where $N (\pt, \yLab )$ is the number of signal candidates extracted from a fit to 
the dimuon invariant mass, \lumi is the integrated luminosity, \branchFrac is the branching fraction for the decay channel resulting in two muons~\cite{ParticleDataGroup:2024cfk}, and \Eff is the 
efficiency associated with the charmonium reconstruction. 
The observable of interest for this analysis is defined as the cross section ratio of \PGyP{2S} to \PJGy mesons for a selection of events containing a multiplicity of charged particles, \nTrackIndex, normalized by the multiplicity-integrated cross section ratio:
\ifthenelse{\boolean{cms@external}}
{
   \begin{multline}
   \text{R}_{\PGyP{2S}/\PJGy}^{\text{Norm.}} \equiv\left(\frac{\sigma_{\PGyP{2S}}}{\sigma_{\PJGy}}\right)_{\nTrackIndex}\bigg/\left(\frac{\sum \sigma_{\PGyP{2S}}}{\sum \sigma_{\PJGy}}\right),
   \end{multline}
}
{
\begin{equation}\label{equ:observable}
   \text{R}_{\PGyP{2S}/\PJGy}^{\text{Norm.}} \equiv\left(\frac{\sigma_{\PGyP{2S}}}{\sigma_{\PJGy}}\right)_{\nTrackIndex}\bigg/\left(\frac{\sum \sigma_{\PGyP{2S}}}{\sum \sigma_{\PJGy}}\right),
\end{equation}
}
where \nTrackIndex indicates a given charged-particle multiplicity selection, and the sum is over all multiplicity bins. In~\cite{LHCb:2023xie}, the same observable is denoted as normalized $\sigma_{\PGyP{2S}}/\sigma_{\PJGy}$. Notably, the factors related to the integrated luminosity, branching fraction, and muon acceptance exactly cancel, and many experimental uncertainties partially cancel in this ratio. Initial-state effects related to nPDFs or other cold nuclear matter effects almost fully cancel as well~\cite{Ferreiro:2014bia}, making this observable a clean probe of potential final-state effects---a theoretically motivated expectation that remains to be validated through experimental confirmation.  

The CMS detector~\cite{CMS:2008xjf,CMS:2023gfb} is a multipurpose, nearly hermetic detector, designed to trigger on~\cite{CMS:2020cmk,CMS:2016ngn} and identify electrons, muons, photons, and hadrons~\cite{CMS:2020uim,CMS:2018rym,CMS:2014pgm}. A global algorithm~\cite{CMS:2017yfk} aims to reconstruct all individual particles in an event, combining information provided by the all-silicon inner tracker and by the crystal electromagnetic, and brass and scintillator hadron calorimeters, operating inside a 3.8\unit{T} superconducting solenoid, with data from the gas-ionization muon detectors embedded in the flux-return yoke outside the solenoid. The hadron forward (HF) calorimeters, made of steel and quartz fibres, extend the pseudorapidity coverage provided by the barrel and endcap detectors.

Events of interest are selected using a two-tiered trigger system. The first level, composed of custom hardware processors, uses information from the calorimeters and muon detectors to select events at a rate of around 100\unit{kHz} within a fixed latency of about 4\mus~\cite{CMS:2020cmk}. The second level, known as the high-level trigger, consists of a farm of processors running a version of the full event reconstruction software optimized for fast processing, and reduces the event rate to around 1\unit{kHz} before data storage~\cite{CMS:2016ngn}. The events used in this analysis were collected using a trigger at first level requiring two muon candidates found in the muon detectors~\cite{CMS:2024qgs}.  

The data were collected in two configurations, in which the directions of the proton and lead beams were reversed. We adopt the convention that the proton-going side of the detector corresponds to positive rapidity values, regardless of the beam directions. For \pPb collisions, the CM reference frame is not at rest with respect to the detector, but rather is boosted in the direction of the outgoing proton beam with a rapidity shift of 
0.465 units. Results are reported versus the charmonia rapidity in the CM reference frame, \yCM, where $\yLab = \yCM + 0.465$.

For the offline analysis, events have to pass a set of selection criteria designed to reject background processes, such as beam-gas collisions and beamscraping events~\cite{CMS:2016xef}. Events are required to have at least one reconstructed primary interaction vertex~\cite{CMS:2014pgm} with a distance from the center of the nominal interaction region of less than 15\unit{cm} along the beam direction and 2\unit{cm} in the plane transverse to the beam direction. To select inelastic hadronic collisions, events are also required to have at least one detector element in each of the HF detectors with energy deposits of more than 3\GeV, which is above the detector noise level. 
In the \pPb data sample, the average number of interactions per bunch crossing, commonly referred to as ``pileup," ranged from 0.10 to 0.25. A procedure similar to that outlined in Refs.~\cite{CMS:2017lrw,CMS:2017kcs,CMS:2013jlh} is used to identify and reject pileup events, effectively removing approximately 3\% of the total events.

Muons of interest are required to pass the ``soft" muon selection and acceptance requirements described in Ref.~\cite{CMS:2024qgs}. The \PJGy and \PGyP{2S} candidates are formed by combining all possible oppositely-charged muon pairs within a single event. 
The rapidity of the resulting candidates in the laboratory reference frame must be $\abs{y_{\text{lab}}} < 2.4$. Additionally, a vertex fit is performed on the muon pair, and there must be a 
probability larger than 1\% that the two muons originate from a common vertex. 

To measure the multiplicity dependence of the cross section ratio, the total reconstructed charged-particle multiplicity (\Ntroff) is calculated in each event using tracks passing the quality selection criteria described in Ref.~\cite{CMS:2017kcs}. In particular, tracks must fall in the kinematic ranges $\abs{\eta^{\text{track}}} < 2.4$ and $\ptTrack>0.4\GeV$. 
The \Ntroff values are corrected for detector inefficiencies and track misidentification rates using Monte Carlo (MC) simulations, with an average correction of around $\sim$21\%. 
The resulting average number of tracks in each \Ntroff range after correction is denoted as \Ntroffcorr. To cancel residual uncertainties related to the \Ntroffcorr calculation, results are presented as a function of the ratio $\Ntroffcorr/\nTrackMB$, where $\nTrackMB = 59.5\pm 0.5$ is the average corrected multiplicity in an inclusive minimum bias (MB) sample. These MB events are selected by requiring energy deposits exceeding approximately 1\GeV in at least one of the two HF calorimeters, and the presence of at least one track with $\ptTrack>0.4\GeV$ in the pixel tracker. The multiplicity selections used in this analysis, when listed in order of increasing multiplicity, correspond to 48.3\%, 18.2\%, 13.6\%, 9.1\%, 5.4\%, and 5.4\% of the MB cross section.

Monte Carlo simulations are used to correct for dimuon reconstruction inefficiency, as well as for parametrizing 
the signal probability density function during signal extraction. Separate MC simulations are used 
to model prompt and nonprompt processes. The simulated events are generated with \PYTHIA8 (v8.212)~\cite{Bierlich:2022pfr} tune CUETP8M1~\cite{CMS:2015wcf} and \EVTGEN (v1.3.0)~\cite{Lange:2001uf} is used to model nonprompt decays. 
The quantum electrodynamics final-state radiation is modeled by \PHOTOS (v3.56)~\cite{Davidson:2010ew} at leading order. Next, 
the \PYTHIA event is embedded in an underlying \pPb collision that is simulated with EPOS LHC (v3400)~\cite{Pierog:2013ria}. 
The events are then processed through a detailed simulation of the CMS detector
using the \GEANTfour~\cite{geant4} package. A separate MC sample is created for each beam direction. 

The MC samples are weighted to match the data distributions that are most relevant for efficiency corrections: those of \Ntroff, the \pt of the charmonium candidates, and the $z$ coordinate of the primary vertex. The individual components of the muon reconstruction efficiency (muon identification, selection, 
and triggering) are measured using single muons from \PJGy meson decays in both simulated and collision data with the ``tag-and-probe" (T\&P) method~\cite{CMS:2011aa,CMS:2024qgs}. The data-to-simulation ratios of single-muon efficiencies obtained from the T\&P method, 
calculated as a function of muon $\pt^{\PGm}$ and pseudorapidity, $\eta^{\PGm}$, are used as weights to fine-tune the dimuon efficiencies obtained from simulation on an event-by-event basis. 

\begin{figure*}[htb]
   \centering
   \includegraphics[width=1.0\textwidth]{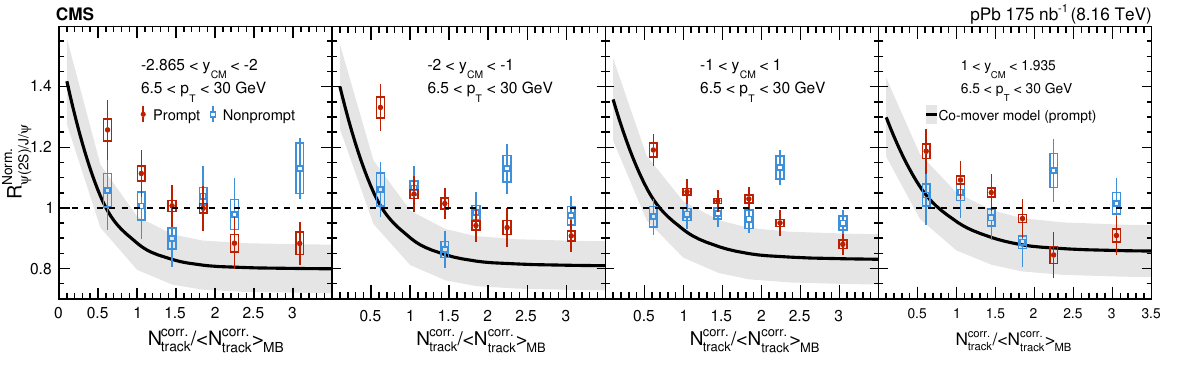}
   \caption{The prompt (red circular points) and nonprompt (blue square points) normalized cross section ratio for $-2.865 < \yCM < -2$ (the leftmost), $-2 < \yCM < -1$ (center-left), $-1 < \yCM < 1$ (center-right), and $1 < \yCM < 1.935$ (the rightmost) in the range $6.5 < \pt < 30 \GeV$. The black line with a gray band shows a model incorporating comover effects~\cite{Ferreiro:2014bia} for prompt charmonia. Error bars represent statistical uncertainties, while boxes display systematic uncertainties.}
   \label{fig:ForwardMid}
\end{figure*}
Candidates are separated into two classes with an enriched fraction of either prompt or nonprompt charmonia, based on whether the dimuon secondary \diMu vertex displacement from the primary collision vertex falls below (for prompt) or above (for nonprompt) a set threshold determined from simulation. For each class, the yields of \PJGy and \PGyP{2S} mesons are calculated by performing an unbinned maximum likelihood fit of the dimuon invariant mass distribution in the range $2.5 < m_{\diMu} < 4.3\GeV$. 
Cross-contamination between the two classes of candidates is corrected using a method similar to that of Refs.~\cite{CMS:2014vjg,CMS:2016wgo,CMS:2018gbb}, with nonprompt contamination estimated to be approximately 5\% for \PJGy and 8\% for \PGyP{2S} candidates, respectively. Details on the prompt and nonprompt vector meson separation, the invariant mass fitting procedure, and examples of invariant mass peaks are provided in \appname~\ref{app:YieldExtraction}.

The systematic uncertainty in the prompt and nonprompt separation procedure is evaluated by calculating the difference between the prompt and nonprompt meson yields with and without correction for residual contamination. It is below 1 and 2\% for the prompt and nonprompt double ratios, respectively. 
The systematic uncertainty associated with the invariant mass fitting procedure includes three components: the choice of the signal function, variations of MC-based fit constraints, and changes of the background function. 
The total systematic uncertainties associated with the fitting procedure on the double ratio remain below 1 and 2\% for the prompt and nonprompt double ratio, respectively, with details of the procedure provided in \appname~\ref{app:YieldExtraction}. 
The statistical uncertainties in the MC sample used to calculate candidate reconstruction efficiencies are propagated to the final result as systematic uncertainties and amount to 0.5\% in the prompt and 1\% in the nonprompt ratio, respectively. 
Finally, the systematic uncertainty due to variations in the pileup filter~\cite{CMS:2017lrw,CMS:2017kcs} amounts to 0.5\%.
Uncertainties associated with the muon trigger and muon identification efficiencies, calculated with the T\&P method, are found to be negligible as a result of the cancellation effect inherent 
in the normalized ratio observable. Similarly, systematic uncertainties related to the variation of the dimuon \pt and \Ntroff spectrum reweighting functions are found to be negligible. Systematic uncertainties from different sources are added in quadrature to estimate the overall systematic uncertainty. When evaluating the systematic uncertainties, variations are made simultaneously
for both \PJGy and \PGyP{2S} candidates to ensure that correlations between the two are appropriately taken into account. The total systematic uncertainty for the inclusive \yCM selection is 1--2 (2--4)\% for prompt (nonprompt) charmonia across all multiplicity ranges studied, and extends up to 4 (8)\% when examining prompt (nonprompt) mesons in the highest multiplicity range and the rapidity range $-2.865 < \yCM < -2$.

The normalized cross section ratios for charmonia in the range $6.5 < \pt < 30 \GeV$ are reported 
in four \yCM ranges as a function of normalized multiplicity in Fig.~\ref{fig:ForwardMid}. The red circular (blue square) points show the results for prompt (nonprompt) charmonia. In all four rapidity ranges examined, a decreasing trend of the prompt cross section ratio is observed, with values falling from around 1.3 at low multiplicity to approximately 0.9 at high multiplicity. This indicates that the charged-particle multiplicity of the event is correlated with an increased tendency to observe prompt charmonia in the ground state compared to excited states. On the other hand, the ratio measurements for nonprompt charmonia do not exhibit any strong multiplicity dependence, and their values are largely consistent with unity. These findings are consistent with previous CMS measurements showing no nuclear modification in b-hadron and b-jet production in \pPb collisions ~\cite{CMS:2015gcq,CMS:2017exb,CMS:2024nrk}. As nonprompt charmonia result from b hadron decays that largely occur after the system has reached a stage of reduced interactions with the surrounding environment, this finding is consistent with the picture that the observed suppression of prompt charmonia arises at an early stage in the evolution of the system.

The data are compared to a prediction calculated using a comover model described in Ref.~\cite{Ferreiro:2014bia}, shown by the solid black line, with the gray band representing the 10\% normalization uncertainty of the model. This uncertainty is fully correlated across all panels in Fig.~\ref{fig:ForwardMid}. The prediction is calculated for prompt charmonia in the same \yCM, \pt, and $\Ntroffcorr/\nTrackMB$ ranges as the data. Comparison of this model with the prompt data reveals reasonable agreement in the rapidity range closest to the proton-going direction ($1 < \yCM < 1.935$), with the data lying on the top edge of the uncertainty band for the model. Although the shape of the distribution seems to be well described in all rapidity ranges, the co-mover model shows an increasing discrepancy with the data as the rapidity selection gets closer to the lead-going direction, i.e., more negative. The discrepancy between data and model is never more than $\sim$2 standard deviations.

\begin{figure}[htb]
   \centering
   \includegraphics[width=0.5\textwidth]{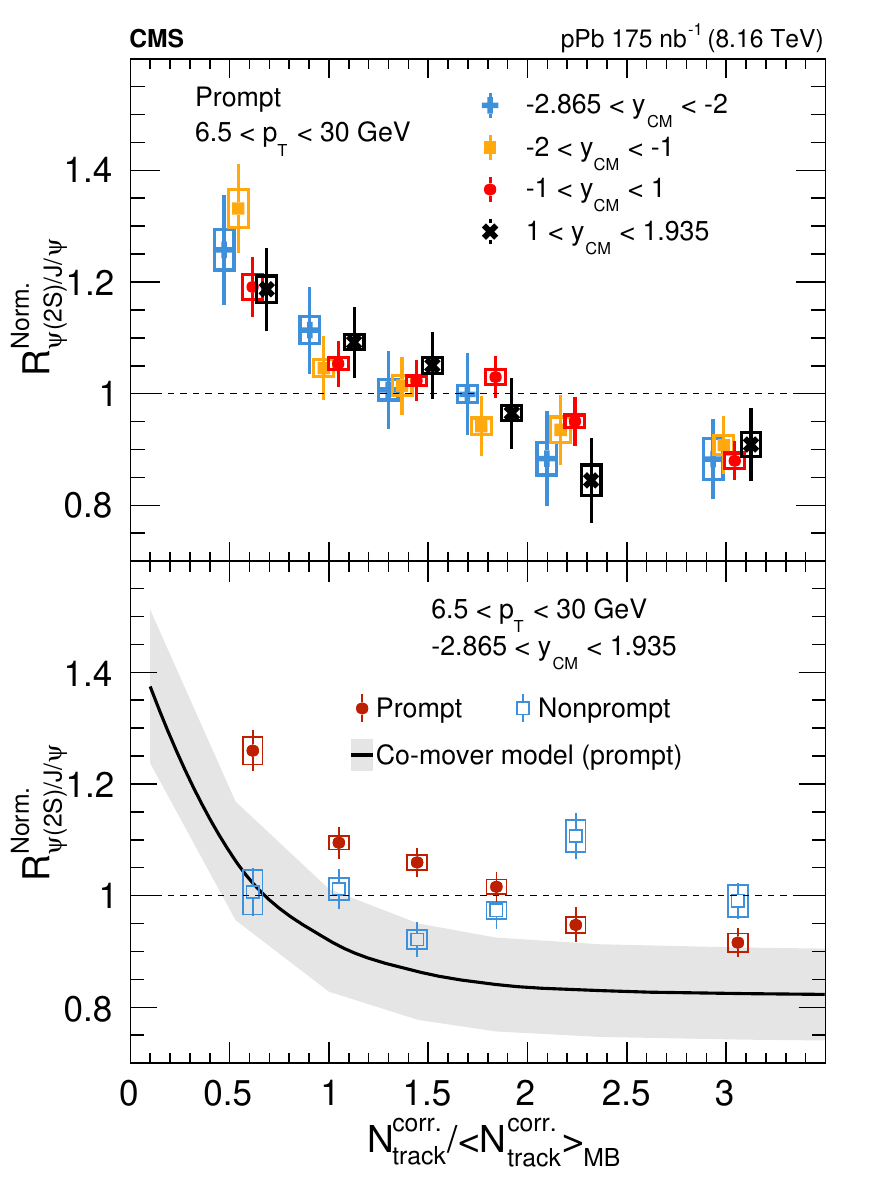}
   \caption{(upper) The prompt normalized cross section ratio versus normalized event charged-particle multiplicity for four different rapidity selections. The different sets of colored data points have a slight horizontal offset applied for visual clarity. (lower) The same observable for prompt (red circular points) and nonprompt (blue square points) mesons when combining all four rapidity selections into a single measurement. The black line with a gray band shows a model incorporating co-mover effects~\cite{Ferreiro:2014bia} for prompt charmonia. Error bars represent statistical uncertainties, while boxes display systematic uncertainties.}
   \label{fig:InclusiveHighpTandMergedpT}
\end{figure}

A comparison between the four different rapidity ranges studied for the prompt data is shown in Fig.~\ref{fig:InclusiveHighpTandMergedpT} (upper). No clear rapidity dependence of the cross section ratio can be seen in the \pt range of 6.5--30\GeV. 
Thus, the rapidity-dependent offset between the data and the co-mover model prediction originates from the model overestimation of the rapidity dependence for this ratio, presumably due to a larger comover density present in the lead-going direction. 
This conclusion appears consistent with previous multiplicity-inclusive measurements of the rapidity dependence of charmonia production in \pPb collisions by the LHCb Collaboration~\cite{LHCb:2024taa}, although larger experimental uncertainties masked the effect. We also note that the lack of rapidity dependence for this observable implies that potential shifts in the effective CM system rapidity, which could arise from correlations between hard process dynamics and event multiplicity~\cite{ATLAS:2014cpa}, appear to be small.

The co-mover model largely focuses on describing charmonium dynamics at relatively low \pt, so it is possible that additional mechanisms related to high \pt charmonium formation, such as jet production, could explain the differences between the data and model.
To explore this possibility, two additional measurements of the same observable in the \pt range 3--6.5\GeV are performed in the rapidity ranges of $-2.865 < \yCM < -2$ and $1 < \yCM < 1.935$ and are detailed in \appname~\ref{app:LowpT}. An indication of a stronger rapidity dependence may be present for this lower \pt selection, but the experimental uncertainties prevent definitive conclusions.

Because of the lack of observed rapidity dependence for high-\pt charmonia, the four rapidity selections are merged into a single measurement, shown in Fig.~\ref{fig:InclusiveHighpTandMergedpT} (lower), to probe the multiplicity dependence of the cross section ratio as precisely as possible. To test a null hypothesis assuming no multiplicity dependence, the agreement between the prompt data and a constant-value fit was evaluated. When accounting for bin-by-bin correlations between data points, the null hypothesis is disfavored with a statistical significance of 5.3 standard deviations. Thus, this measurement constitutes the first observation of multiplicity dependence of the cross section ratio for charmonia produced in \pA collisions. The null hypothesis is only disfavored by 2.1 standard deviations when the same procedure is applied to the nonprompt data.

Given that initial-state effects almost completely cancel in the normalized cross section ratio, the prompt measurements provide a clear indication of the presence of multiplicity-dependent final-state modifications to excited charmonium production. Furthermore, the lack of multiplicity dependence for the nonprompt data indicates that this final-state effect occurs relatively early in the evolution of the system. The comover effect is one such mechanism that naturally explains these properties of the cross section ratio, and it appears to qualitatively describe the data well~\cite{Ferreiro:2014bia}. However, other explanations, e.g., the formation of a small amount of QGP in \pPb collisions, are also possible. Future comparisons with models incorporating QGP effects, e.g., Ref.~\cite{Du:2018wsj}, could test if this alternative mechanism produces the correct multiplicity and rapidity dependence of the cross section ratio, potentially yielding insights into which of these two final-state mechanisms is more likely to occur. 
Tabulated results are provided in the HEPData record for this analysis~\cite{hepdata}.

In summary, this Letter presents a measurement of the normalized production cross section ratio of \PGyP{2S} over \PJGy mesons as a function of
charged-particle multiplicity in proton-lead collisions at a center-of-mass energy per nucleon pair of
8.16\TeV with a data sample collected with the CMS detector. The ratio is 
measured for both prompt and nonprompt \PGyP{2S} and \PJGy mesons separately. For the first time in proton-ion collisions, a statistically significant multiplicity-dependent 
modification of the ratio is observed for prompt charmonia, indicating a decreasing trend with increasing multiplicity. 
The ratio of the nonprompt cross sections shows no dependency on the multiplicity. 
No appreciable rapidity dependence is observed within the charmonium rapidity and transverse momentum ranges probed. These observations support the presence of final state effects dissociating weakly bound excited states of charmonia more than the ground state, which can be interpreted as co-mover dissociation or, potentially, the formation of quark-gluon plasma. These new data provide powerful tests for hadronization models in small collision systems.

\begin{acknowledgments}
We congratulate our colleagues in the CERN accelerator departments for the excellent performance of the LHC and thank the technical and administrative staffs at CERN and at other CMS institutes for their contributions to the success of the CMS effort. In addition, we gratefully acknowledge the computing centers and personnel of the Worldwide LHC Computing Grid and other centers for delivering so effectively the computing infrastructure essential to our analyses. Finally, we acknowledge the enduring support for the construction and operation of the LHC, the CMS detector, and the supporting computing infrastructure provided by the following funding agencies: SC (Armenia), BMBWF and FWF (Austria); FNRS and FWO (Belgium); CNPq, CAPES, FAPERJ, FAPERGS, and FAPESP (Brazil); MES and BNSF (Bulgaria); CERN; CAS, MoST, and NSFC (China); MINCIENCIAS (Colombia); MSES and CSF (Croatia); RIF (Cyprus); SENESCYT (Ecuador); ERC PRG, RVTT3 and MoER TK202 (Estonia); Academy of Finland, MEC, and HIP (Finland); CEA and CNRS/IN2P3 (France); SRNSF (Georgia); BMBF, DFG, and HGF (Germany); GSRI (Greece); NKFIH (Hungary); DAE and DST (India); IPM (Iran); SFI (Ireland); INFN (Italy); MSIP and NRF (Republic of Korea); MES (Latvia); LMTLT (Lithuania); MOE and UM (Malaysia); BUAP, CINVESTAV, CONACYT, LNS, SEP, and UASLP-FAI (Mexico); MOS (Montenegro); MBIE (New Zealand); PAEC (Pakistan); MES and NSC (Poland); FCT (Portugal); MESTD (Serbia); MICIU/AEI and PCTI (Spain); MOSTR (Sri Lanka); Swiss Funding Agencies (Switzerland); MST (Taipei); MHESI and NSTDA (Thailand); TUBITAK and TENMAK (Turkey); NASU (Ukraine); STFC (United Kingdom); DOE and NSF (USA).
\end{acknowledgments}

\bibliography{auto_generated}
\clearpage
\appendix

\section{Yield extraction\label{app:YieldExtraction}}

\begin{figure}[htb]
   \centering
   \includegraphics[width=0.45\textwidth]{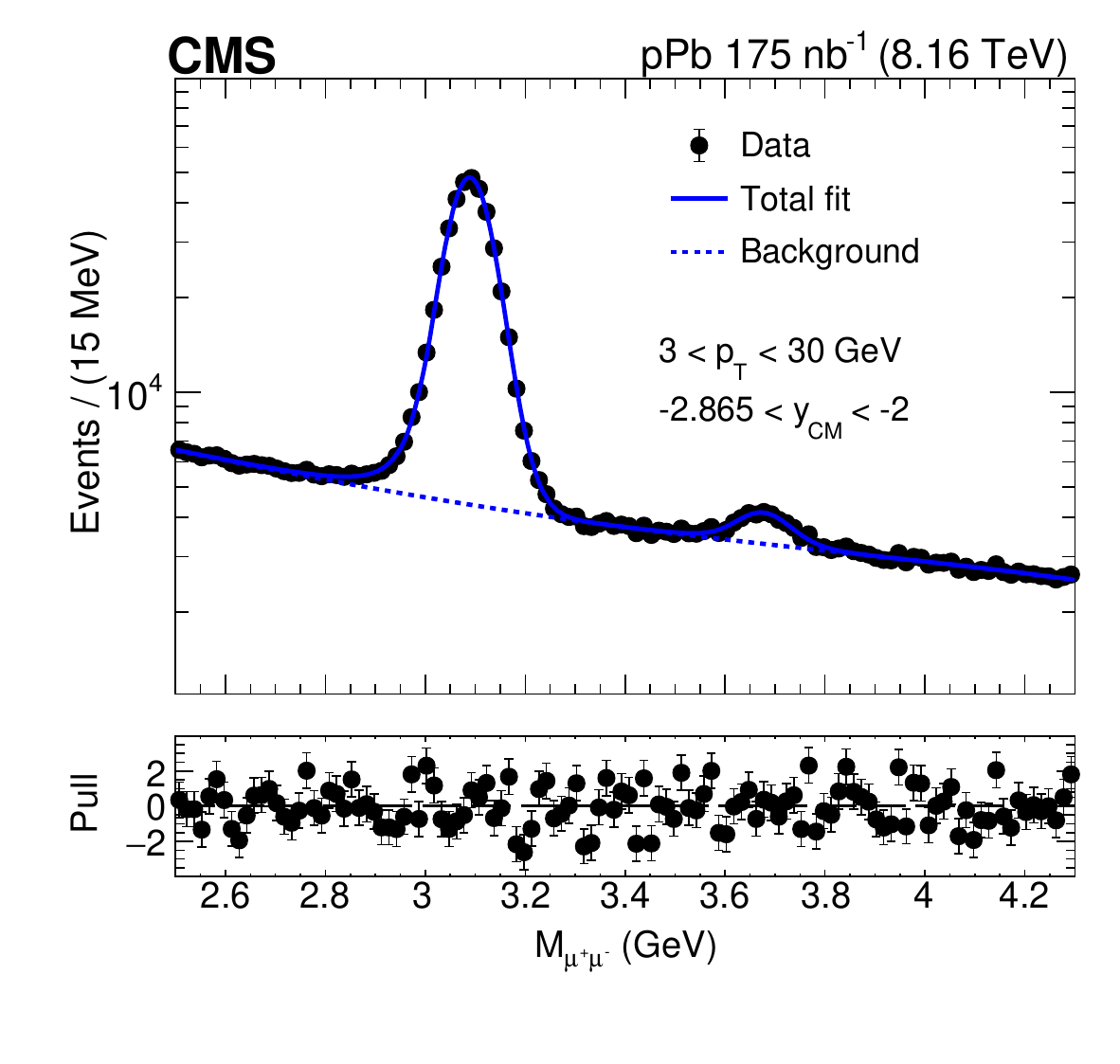}
   \includegraphics[width=0.45\textwidth]{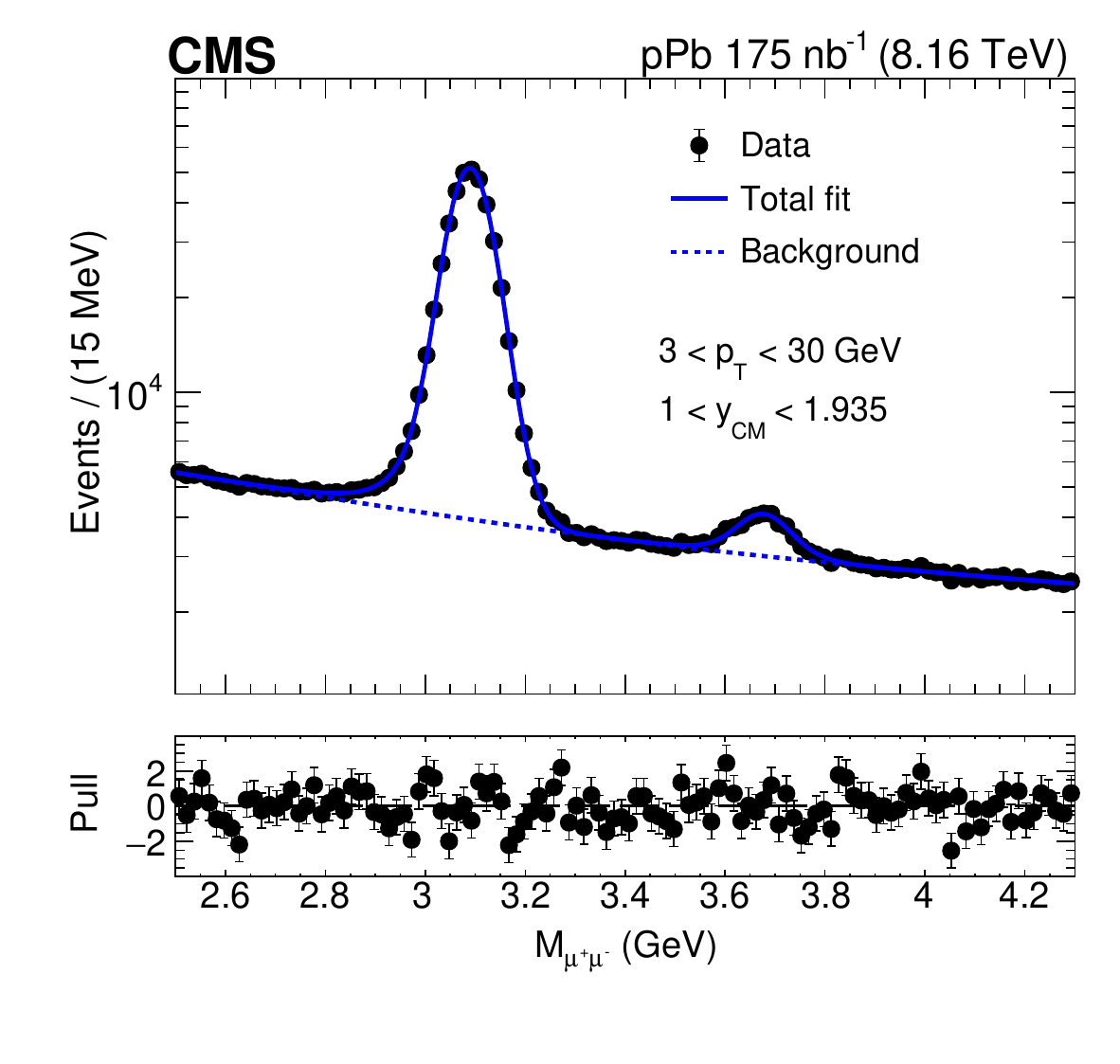}
   \caption{Dimuon invariant mass distribution showing the \PJGy and \PGyP{2S} peaks in \pPb data, for $-2.865 < \yCM < -2$ (\cmsLeft) and $1 < \yCM< 1.935$ (\cmsRight) for $3 < \pt < 30\GeV$. The total fit is shown with a solid blue line, and the dotted blue line represents the background contribution.}
   \label{fig:MassFit}
\end{figure} 

The invariant mass fits are performed by maximizing an unbinned and extended likelihood in the region of $2.5 < m_{\diMu} < 4.3 \GeV$ 
using a method that is similar to that of Ref.~\cite{CMS:2018gbb}.
Both the \PJGy and \PGyP{2S} nominal signal shapes are described using the sum of two Crystal Ball (CB)~\cite{CrystalBallRef} functions. The background distribution is described by a Chebyshev polynomial of order $(0 \leq N \leq 3)$, with the appropriate order for each analysis bin selected based on log-likelihood ratio tests.
The tail parameters are fixed to the values calculated using a \PJGy MC sample, and both the \PJGy and \PGyP{2S}
signal functions use the same values of these tail parameters. In the fits to the data, the ratio of the widths of two CB functions and 
the relative yield fractions of two CB functions are also fixed to the values found in a \PJGy MC sample. 
The mean and width of the \PGyP{2S} invariant mass distribution are obtained by fitting the mean and width of the \PJGy mass peak and scaling it by the ratio of the \PGyP{2S} to \PJGy masses, as calculated using world-average values~\cite{ParticleDataGroup:2024cfk}. Representative examples of invariant mass distributions for two different rapidity selections can be seen in Fig.~\ref{fig:MassFit}. 
To evaluate the systematic uncertainty associated with the choice of signal probability density function, a CB function plus a Gaussian function is used instead of the nominal sum of two CB functions. 
For the systematic component related to the background model similar to that of Ref.~\cite{CMS:2018gbb}, the order of the Chebyshev polynomial is varied by increasing $N$ to $N+1$ if the nominal fit uses $N \leq 2$, and decreasing to $N-1$ for $N>2$. 
Lastly, the uncertainty associated with constraining the signal shape parameters to those obtained from simulations is assessed by individually releasing each parameter and repeating the fit. The differences between the nominal double ratio values and those obtained when a parameter is allowed to vary are calculated, and the root-mean-square of these differences is taken as the systematic uncertainty.

Prompt and nonprompt charmonia candidates are separated using a method that is similar to those used in earlier analyses such as Refs.~\cite{CMS:2014vjg,CMS:2016wgo,CMS:2018gbb}. The separation is done before performing the invariant mass fit by selecting on the pseudoproper decay lengths of the candidates. The decay length is expressed as $\lDecay = \Lxyz m/\abs{p}$, where \Lxyz is the distance between the primary and the dimuon vertices in the laboratory frame, $\abs{p}$ is the magnitude of the dimuon momentum, and $m$ is the world-average \PJGy or \PGyP{2S} mass~\cite{ParticleDataGroup:2024cfk}, depending on the meson being studied. Prompt (nonprompt) charmonia are measured by rejecting dimuons with \lDecay above (below) a threshold value that is determined by examining the efficiency of selecting prompt (nonprompt) charmonia, denoted as \PromptEff (\NonPromptEff), in simulated samples. The threshold is chosen separately for each rapidity selection so that the efficiency for prompt \PJGy is fixed at 90\%. This value is around 35--50\mum for charmonia candidates in the \pt range 6.5--30\GeV and corresponds to a nonprompt charmonium
rejection rate of ${>}86$\%. In the \pt range of 3.0--6.5 \GeV, the threshold value is in the range of 74--78\mum, with a nonprompt charmonium rejection rate of ${>}81$\%. The \lDecay cut that is used for \PJGy is also used for \PGyP{2S} candidates, resulting in a prompt \PGyP{2S} efficiency that is slightly ${>}90$\%. Residual contamination of the prompt or nonprompt charmonium samples by candidates of the other type is corrected after the invariant mass fit performed. The corrected prompt meson yield can be expressed as 
\begin{equation*}
\NPrompt = \frac{(1-\NonPromptEff)\NPass-\NonPromptEff\NFail}{\PromptEff-\NonPromptEff},
\end{equation*}
where \NPass (\NFail) is the number of dimuon candidates in data below (above) the \lDecay selection. The corrected nonprompt meson yield is given by the difference between the total number of candidates and the prompt candidates, $\NNonPrompt = (\NPass + \NFail) - \NPrompt$.

\section{Results in low-\texorpdfstring{\pt}{pt} region\label{app:LowpT}}
\begin{figure}[htbp!]
   \centering
   \includegraphics[width=0.5\textwidth]{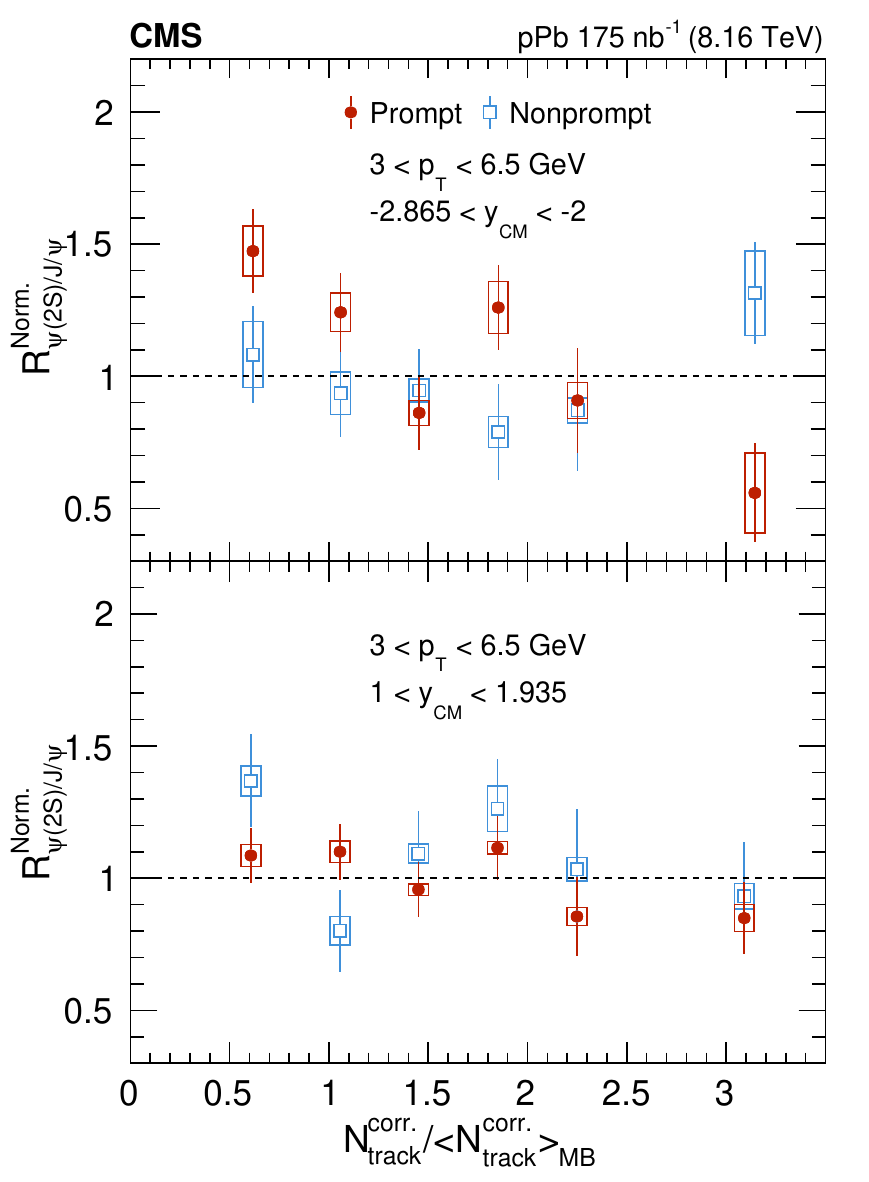}
   \caption{The prompt (red circular points) and nonprompt (blue square points) normalized cross section ratio for $-2.865 < \yCM < -2$ (upper), and $1 < \yCM < 1.935$ (lower) in the range $3 < \pt < 6.5\GeV$. Error bars represent statistical uncertainties, while boxes display systematic uncertainties.}
   \label{fig:LowpTFig}
\end{figure}

Because of the larger backgrounds in the low-\pt region, which reduces the \PGyP{2S} signal significance, as well as for muon acceptance considerations,
measurements in the range $3 < \pt < 6.5\GeV$ are limited to the endcap regions of the CMS 
detector, i.e., the rapidity regions of $-2.865 < \yCM < -2$ and $1 < \yCM < 1.935$. Measurements of the normalized cross section ratio in this kinematic region are presented in Fig.~\ref{fig:LowpTFig}. A decreasing trend appears to be present for the prompt charmonia in the lead-going rapidity selection, while a relatively flat trend is present for charmonia traveling in the proton-going direction. However, the uncertainties in this \pt region are also larger than those for the higher-\pt selection because of lower reconstruction efficiencies and signal significances. When evaluating the agreement of the prompt data with a constant-value fit, the null hypothesis of no multiplicity dependence is only disfavored with a statistical significance of 1.7 (0.6) standard deviations in the lead- (proton-)going direction. Thus, there may be a hint of rapidity-dependence of the normalized cross section ratio for this lower \pt selection, but measurement uncertainties prevent definitive conclusions at this point. Like the measurements performed at higher \pt, the nonprompt ratio values are largely consistent with unity.

\cleardoublepage \section{The CMS Collaboration \label{app:collab}}\begin{sloppypar}\hyphenpenalty=5000\widowpenalty=500\clubpenalty=5000\input{HIN-24-001-public-authorlist.tex}\end{sloppypar}
%%% END EDITABLE REGION %%%
% skeleton_end
\end{document}

%% file: HIN-24-001-public-authorlist.tex
\cmsinstitute{Yerevan Physics Institute, Yerevan, Armenia}
{\tolerance=6000
V.~Chekhovsky, A.~Hayrapetyan, V.~Makarenko\cmsorcid{0000-0002-8406-8605}, A.~Tumasyan\cmsAuthorMark{1}\cmsorcid{0009-0000-0684-6742}
\par}
\cmsinstitute{Institut f\"{u}r Hochenergiephysik, Vienna, Austria}
{\tolerance=6000
W.~Adam\cmsorcid{0000-0001-9099-4341}, J.W.~Andrejkovic, L.~Benato\cmsorcid{0000-0001-5135-7489}, T.~Bergauer\cmsorcid{0000-0002-5786-0293}, S.~Chatterjee\cmsorcid{0000-0003-2660-0349}, K.~Damanakis\cmsorcid{0000-0001-5389-2872}, M.~Dragicevic\cmsorcid{0000-0003-1967-6783}, P.S.~Hussain\cmsorcid{0000-0002-4825-5278}, M.~Jeitler\cmsAuthorMark{2}\cmsorcid{0000-0002-5141-9560}, N.~Krammer\cmsorcid{0000-0002-0548-0985}, A.~Li\cmsorcid{0000-0002-4547-116X}, D.~Liko\cmsorcid{0000-0002-3380-473X}, I.~Mikulec\cmsorcid{0000-0003-0385-2746}, J.~Schieck\cmsAuthorMark{2}\cmsorcid{0000-0002-1058-8093}, R.~Sch\"{o}fbeck\cmsAuthorMark{2}\cmsorcid{0000-0002-2332-8784}, D.~Schwarz\cmsorcid{0000-0002-3821-7331}, M.~Sonawane\cmsorcid{0000-0003-0510-7010}, W.~Waltenberger\cmsorcid{0000-0002-6215-7228}, C.-E.~Wulz\cmsAuthorMark{2}\cmsorcid{0000-0001-9226-5812}
% ...
\par}
\vskip\cmsinstskip
\dag:~Deceased\\
$^{1}$Also at Yerevan State University, Yerevan, Armenia\\
$^{2}$Also at TU Wien, Vienna, Austria\\
% ...
